\documentclass[sigconf, nonacm]{acmart}

\AtBeginDocument{%
  }

\setcopyright{acmlicensed}
\copyrightyear{2018}
\acmYear{2018}
\acmDOI{XXXXXXX.XXXXXXX}

\usepackage{amsmath} 
\usepackage{algorithm} 
\usepackage{algorithmic} 
\usepackage{graphicx} 
\usepackage{booktabs} 
\usepackage{float} 

\usepackage[table]{xcolor}
\definecolor{mygray}{gray}{0.9} 

\begin{document}

\title{Deep GraphRAG: A Balanced Approach to Hierarchical Retrieval and Adaptive Integration}

\author{Yuejie Li}
\email{liyuejie.lyj@antgroup.com}
\affiliation{%
  \institution{Ant Group}
  \country{China}
}

\author{Ke Yang}
\email{zhulang.yk@antgroup.com}
\affiliation{%
  \institution{Ant Group}
  \country{China}
}

\author{Tao Wang}
\email{wangtop@zju.edu.cn}
\affiliation{%
  \institution{Zhejiang University}
  \country{China}
}

\author{Bolin Chen}
\email{bolin.cbl@antgroup.com}
\affiliation{%
  \institution{Ant Group}
  \country{China}
}

\author{Bowen Li}
\email{zhikong.lbw@antgroup.com}
\affiliation{%
  \institution{Ant Group}
  \country{China}
}

\author{Chengjun Mao}
\email{chengjun.mcj@antgroup.com}
\affiliation{%
  \institution{Ant Group}
  \country{China}
}

\renewcommand{\shortauthors}{Li et al.}

\begin{abstract}
Graph-based Retrieval-Augmented Generation (GraphRAG) frameworks face a trade-off between the comprehensiveness of global search and the efficiency of local search. Existing methods are often challenged by navigating large-scale hierarchical graphs, optimizing retrieval paths, and balancing exploration-exploitation dynamics, frequently lacking robust multi-stage re-ranking. To overcome these deficits, we propose Deep GraphRAG, a framework designed for a balanced approach to hierarchical retrieval and adaptive integration. It introduces a hierarchical global-to-local retrieval strategy that integrates macroscopic inter-community and microscopic intra-community contextual relations. This strategy employs a three-stage process: (1) inter-community filtering, which prunes the search space using local context; (2) community-level refinement, which prioritizes relevant subgraphs via entity-interaction analysis; and (3) entity-level fine-grained search within target communities. A beam search-optimized dynamic re-ranking module guides this process, continuously filtering candidates to balance efficiency and global comprehensiveness. Deep GraphRAG also features a Knowledge Integration Module leveraging a compact LLM, trained with Dynamic Weighting Reward GRPO (DW-GRPO). This novel reinforcement learning approach dynamically adjusts reward weights to balance three key objectives: relevance, faithfulness, and conciseness. This training enables compact models (1.5B) to approach the performance of large models (70B) in the integration task. Evaluations on Natural Questions and HotpotQA demonstrate that Deep GraphRAG significantly outperforms baseline graph retrieval methods in both accuracy and efficiency.
\end{abstract}


\maketitle

\section{Introduction}

\begin{figure*}[t]
\centering
\includegraphics[width=2\columnwidth]{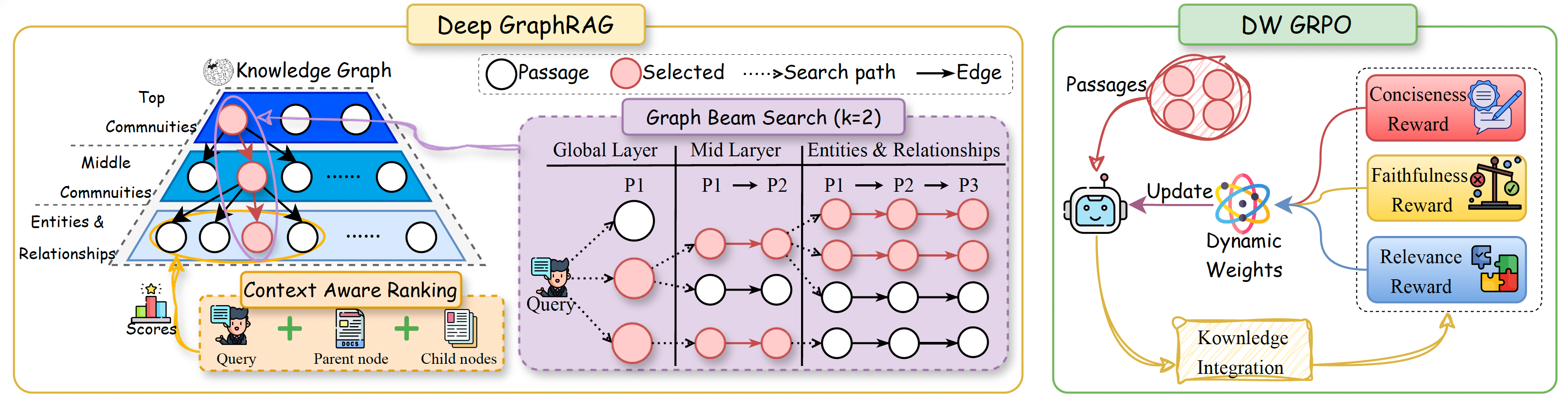}
\caption{Deep GraphRAG framework overview. The retrieval module uses Graph Beam Search and Context Aware Ranking on a hierarchical Knowledge Graph. The knowledge integration module employs Dynamic Weighting Reward GRPO (DW GRPO).}
\label{fig1}
\end{figure*}

While Retrieval-Augmented Generation (RAG) effectively mitigates common LLM challenges such as hallucination and knowledge cutoff \cite{lewis2020retrieval}, conventional vector-based retrieval methods demonstrate limitations in complex reasoning tasks that necessitate structural comprehension \cite{2401_18059,2305_18703,2406_04369,2410_08815,2207_06300,2310_04408,2412_01572}. This deficit has catalyzed the development of knowledge graph-based RAG (GraphRAG) \cite{Edge2024}. However, even advanced approaches—including GNN-enhanced frameworks \cite{luo2025gfmraggraphfoundationmodel}, modular indexing systems \cite{li2025rglgraphcentricmodularframework}, agent-based graph readers \cite{li2024graphreaderbuildinggraphbasedagent}, and neurobiologically inspired retrieval \cite{gutiérrez2025hipporagneurobiologicallyinspiredlongterm}—exhibit notable limitations in graph traversal.

Specifically, these methods inadequately resolve the exploration-exploitation tradeoff. Coarse-grained community summarization (e.g., Map-Reduce) often sacrifices fine-grained contextual relevance. Concurrently, the absence of multi-stage re-ranking mechanisms can lead to local optima trapping and disconnection between different levels of graph abstraction \cite{Han2024, Microsoft2024}.
 
To address these limitations, we propose Deep GraphRAG, a hierarchical retrieval framework featuring a dynamic, beam-search-inspired re-ranker. As depicted in Figure \ref{fig1}, this framework systematically integrates:
(1) Inter-community filtering to prune the search space via macroscopic topology;
(2) Community-level refinement utilizing entity-interaction graph analysis;
(3) Fine-grained entity retrieval coupled with contextual re-ranking.
This tri-stage architecture is designed to balance global structural awareness with local semantic precision, thereby addressing the limitations of prior models.

Moreover, reinforcement learning algorithms such as TRPO \cite{schulman2015trust}, DPO \cite{rafailov2023direct}, PPO \cite{schulman2017proximal} and GRPO \cite{shao2024deepseekmath} employ fixed weights across multiple reward signals, failing to account for the dynamic trade-offs that may arise during optimization. To address this limitation, we propose Dynamic Weighting Reward GRPO (DW-GRPO), a novel framework that adaptively learns and updates reward coefficients during policy optimization. Empirical results demonstrate that DW-GRPO significantly improves the performance of a compact 1.5B LLM on knowledge integration tasks.

\section{Methods}
\subsection{Deep GraphRAG}
We propose Deep GraphRAG, a retrieval framework using a graph's hierarchical community structure to enhance accuracy via a multi-stage, top-down search.

\subsubsection{Graph Construction and Hierarchy}
We construct the base graph $G=(V, E)$ from the corpus $K$ using a rigorous three-step pipeline designed to maximize structural integrity and semantic retention:

\textbf{Text Chunking and Extraction:} We segment the corpus $K$ using a sliding window approach with a fixed size of $T=600$ tokens and an overlap of $O=100$ tokens to mitigate boundary information loss. For each chunk, we deploy Qwen2.5-72B-Instruct (temperature set to 0 to ensure deterministic output) with a specialized prompt to extract entities and directed relationships. Unlike standard triple-based extraction, we enforce the generation of concise natural language descriptions for every edge to capture complex semantic nuances.

\textbf{Entity Resolution:} To maintain topological consistency, we implement a hybrid resolution strategy. Initial candidate pairs for merging are identified by computing cosine similarity on entity descriptions embedded via the \texttt{bge-m3}\cite{bge_m3} model. Pairs exceeding a strict similarity threshold ($\tau > 0.95$) undergo a secondary verification step where an LLM acts as a discriminator to confirm if the nodes refer to the same real-world concept (e.g., merging "U.S." and "United States").

\textbf{Hierarchy Generation:} Upon the resolved base graph $G$, we construct a multi-granular 3-level community hierarchy $\mathcal{C}$. This is achieved by recursively applying the weighted Louvain algorithm with a standard resolution parameter $\gamma=1.0$. This bottom-up process partitions the graph into increasingly coarse semantic clusters, creating a tree structure where level $L=0$ comprises individual entities and levels $L>0$ represent abstract community summaries.

We then generate context-aware representations $D(\cdot)$ for all nodes and communities using a pre-trained sentence-embedding model. Community representations $D(c)$ are derived by mean pooling their sub-community vectors (Eq. \ref{eq:community-agg}), while node representations $D(v)$ concatenate local descriptions with their parent's representation $D_{parent}(\cdot)$.
\begin{equation}
D_{\text{sub}}(c) = \frac{1}{|C_{\text{sub}}(c)|} \sum_{c' \in C_{\text{sub}}(c)} D(c')
\label{eq:community-agg}
\end{equation}

\subsubsection{Retrieval Process}
The retrieval (Algorithm \ref{alg:overall-process}) employs a coarse-to-fine beam search ($k=3$) to traverse $\mathcal{C}$ from top to bottom. To balance efficiency and accuracy, we adopt a hybrid scoring strategy: relevance at the top coarse level is quickly estimated using re-ranker, while the intermediate expansion utilizes a \textbf{bge-reranker-v2-m3} to capture subtle semantic alignments between the query $q$ and community summaries. 

At the final entity level, we ensure topological awareness by constructing \textbf{context-aware representations}. Specifically, the representation of a candidate entity $v$ is dynamically formed by concatenating its local embedding with its parent community's vector (i.e., $D_{ctx}(v) = [D(v); D(c_{parent})]$) before final scoring.

\begin{algorithm}[t]
\caption{Deep GraphRAG Process}
\label{alg:overall-process}
\textbf{Input}: Query $q$, Community $C$, Beam width $k$, Top-$m$ entities $m$ \\
\textbf{Output}: Final result set $R$
\begin{algorithmic}[1]
\STATE \textbf{Phase 1: Top-level Search} (Coarse Filtering)
\STATE \hspace{1em} Score $C_{top}$ via $\text{Re-ranker}(q, D(c_{top}))$
\STATE \hspace{1em} $C_{mid} \leftarrow \text{Top-k results from } C_{top}$

\STATE \textbf{Phase 2: Middle-level Search} (Fine-grained Re-ranking)
\STATE \hspace{1em} Expand $C_{mid}$ to sub-communities $C'_{mid}$
\STATE \hspace{1em} Score $C'_{mid}$ via $\text{Re-ranker}(q, D(c_{mid})+D(c'_{mid}))$ 
\STATE \hspace{1em} $C_{stop}, V_{cand} \leftarrow \text{Top-k results from } C'_{mid}$

\STATE \textbf{Phase 3: Entity-level Search} (Context-Aware)
\STATE \hspace{1em} Expand $C_{stop}$ to candidate entities $V_{cand}$
\FOR{each entity $v \in V_{cand}$ with parent $c_{parent}$}
    \STATE \hspace{1em} $D_{ctx}(v) \leftarrow \text{Concat}(D(v), D(c_{parent}))$
    \STATE \hspace{1em} $Score(v) \leftarrow \text{sim}_{\cos}(q, D_{ctx}(v))$
\ENDFOR
\STATE \hspace{1em} $R_{pre} \leftarrow \text{Top-m results from } V_{cand}$ based on $Score(v)$

\STATE \textbf{Phase 4: Knowledge Integration}
\STATE \hspace{1em} $R \leftarrow \text{Hierarchical-Integration}(R_{pre})$
\STATE \textbf{return} $R$
\end{algorithmic}
\end{algorithm}

\subsection{Dynamic Weighting Reward GRPO (DW-GRPO)}
\subsubsection{DW-GRPO in Deep GraphRAG}
In the Deep GraphRAG framework, knowledge integration generates distilled knowledge $C$. $C$ is a critical input for subsequent knowledge extraction based on user queries $Q$. The module aims to filter redundancy and suppress hallucinations, emphasizing three core objectives.
We define three rewards for optimizing the distilled knowledge $C$ relative to the original text $K$ and query $Q$:

\begin{enumerate}
\item \textbf{Relevance ($r_{\text{rel}}$):} Measures how well $C$ answers $Q$. We use a pre-trained cross-encoder model (bge-reranker-v2-m3) $f_{\text{cross}}$ to score the query-output pair:
    \begin{equation}
    r_{\text{rel}} = f_{\text{cross}}(Q, C)
    \end{equation}
\item \textbf{Faithfulness ($r_{\text{faith}}$):} Measures the semantic fidelity of $C$ to the original knowledge $K$. We use the F1-score from BERTScore (bge-m3) $f_{\text{BERT}}$:
    \begin{equation}
    r_{\text{faith}} = f_{\text{BERT}}(C, K)
    \end{equation}
\item \textbf{Conciseness ($r_{\text{conc}}$):} Penalizes verbosity to encourage succinct summaries. We define this as a normalized length-based reward:
    \begin{equation}
    r_{\text{conc}} = \max\left(0, 1 - \frac{\text{len}(C)}{\text{len}(K)}\right)
    \end{equation}
\end{enumerate}
These three rewards ($r_1, r_2, r_3$) form the basis for the dynamic weighting mechanism.

\subsubsection{Dynamic Weighting Reward-GRPO}
A limitation of GRPO in multi-reward settings is its use of static weights, which leads to suboptimal performance (the seesaw effect \cite{10.1145/3383313.3412236}). We propose **Dynamic Weighting Reward GRPO (DW-GRPO)**, which replaces static weights with a policy-aware adaptive weighting mechanism.

We incorporate a dynamic parameter $w$ to adjust rewards: $\widetilde{r}=\sum_j{w_jr_j}$. The estimated advantage $\widetilde{\hat{A}}$ is computed as:
\begin{align}
\widetilde{\hat{A}} = \sum_j{w_jr_j} - \frac{\sum_j{w_jr_j} - \text{mean}(\sum_j{w_jr_j})}{\text{std}(\sum_j{w_jr_j})}
\end{align}

The core mechanism of DW-GRPO is to dynamically allocate higher weights to reward components that demonstrate slower growth rates, maximizing long-term returns. Let $\Delta r_j = \max(\mathbf{r}_j^{(t,\tau)}) - \min(\mathbf{r}_j^{(t,\tau)})$ be the reward range over window $\tau$. We fit a linear model to estimate the reward's rate of change ${\text{slope}_j}$.

The normalized rate of change $ \alpha_{j}(t-1) $ is:
\begin{align}
\alpha_{j}(t-1) =
\begin{cases}
0, & \text{if } \Delta r_j = 0 \\
\frac{\text{slope}_j}{\Delta r_j}, & \text{otherwise}
\end{cases}
\end{align}
where the slope is determined by least-squares fitting:
\[
\text{slope}_j = \arg\min_{k, b} \|\mathbf{r}_j^{(t,w)} - (k \mathbf{x} + b)\|^2.
\]
The dynamic weights $w_{j}(t)$ are then derived using the softmax function with temperature $T$, where $W = \sum_j w_{j,0}$ maintains the total scale:
\begin{align}
w_{j}(t) = \frac{W\exp\left(-1 \cdot \alpha_{j}(t-1)/T\right)}{\sum_j \exp\left(-1 \cdot \alpha_{j}(t-1)/T\right)}
\end{align}

\section{Experimental Setup}

\subsection{Datasets}
We evaluated the system on two distinct datasets, Natural Questions (NQ) \cite{10.1162/tacl_a_00276} and HotpotQA \cite{yang2018hotpotqadatasetdiverseexplainable}, to assess its generalization ability and domain adaptability.

\subsection{Question Categorization and Baselines}
To evaluate performance on problems of varying complexity, we categorized test questions into three types based on the structure of their ground-truth answer paths in the knowledge graph:
\begin{itemize}
    \item \textbf{Local Questions (LQ):} Answerable by focusing on 1-2 directly connected entity nodes. Tests retrieval of specific facts.
    \item \textbf{Global Questions (GQ):} Require reasoning across more than 2 entities, often spanning different communities. Tests summarization or comparison.
    \item \textbf{Comprehensive Questions (CQ):} Require a combination of specific local facts (like LQ) and broader aggregated context (like GQ).
\end{itemize}
This objective categorization allows for a fine-grained analysis of each method's strengths.

We compare Deep GraphRAG against three baseline retrieval methods: \textbf{Local Search (LS)} (standard dense vector retrieval over all entity nodes), Microsoft's \textbf{Global Search (GS)} (a map-reduce summarization strategy), and \textbf{DRIFT Search (DS)} (a recursive search method) \cite{Microsoft2024}.

\begin{table*}[t]
\centering
\caption{Performance Comparison of Different Models (NQ and HotpotQA). Best total EM in bold.}
\label{tab:performance_combined}
\resizebox{1.0\textwidth}{!}{%
\begin{tabular}{l|l|l|r|r|r|r}
\toprule
Method & Knowledge Integration & Generation & \multicolumn{1}{c|}{\begin{tabular}[c]{@{}c@{}}EM-LQ (\%)\end{tabular}} & \multicolumn{1}{c|}{\begin{tabular}[c]{@{}c@{}}EM-GQ (\%)\end{tabular}} & \multicolumn{1}{c|}{\begin{tabular}[c]{@{}c@{}}EM-CQ (\%)\end{tabular}} & \multicolumn{1}{c}{\begin{tabular}[c]{@{}c@{}}EM-Total (\%)\end{tabular}} \\
\midrule
\multicolumn{7}{l}{\textbf{NQ Dataset}} \\
\midrule
Local Search & & Qwen2.5 72B & 41.32 & 16.58 & \textbf{16.10} & 30.13 \\
Global Search & Qwen2.5 72B & Qwen2.5 72B & 20.48 & 31.19 & 0.10 & 22.63 \\
Drift Search & Qwen2.5 72B & Qwen2.5 72B & 42.43 & 54.15 & 14.70 & 42.78 \\
\rowcolor{mygray}
Deep GraphRAG & Qwen2.5 72B & Qwen2.5 72B & \textbf{45.36} & \textbf{55.08} & 14.70 & \textbf{44.69} \\
\rowcolor{mygray}
Deep GraphRAG & Qwen2.5 1.5B & Qwen2.5 72B & 24.00 & 28.00 & 3.00 & 21.64 \\
\rowcolor{mygray}
Deep GraphRAG & Qwen2.5 1.5B-DW GRPO & Qwen2.5 72B & 44.80 & 54.00 & 13.00 & 42.36 \\
\midrule
Local Search & & DeepSeek-R1 & 41.09 & 18.04 & \textbf{23.20} & 31.36 \\
Global Search & DeepSeek-R1 & DeepSeek-R1 & 20.52 & 32.15 & 16.40 & 23.79 \\
Drift Search & Qwen2.5 72B & DeepSeek-R1 & 45.23 & \textbf{53.42} & 11.00 & 43.61 \\
\rowcolor{mygray}
Deep GraphRAG & Qwen2.5 72B & DeepSeek-R1 & \textbf{45.36} & 51.00 & 19.60 & \textbf{43.98} \\
\rowcolor{mygray}
Deep GraphRAG & Qwen2.5 1.5B & DeepSeek-R1 & 24.40 & 29.00 & 3.50 & 22.27 \\
\rowcolor{mygray}
Deep GraphRAG & Qwen2.5 1.5B-DW GRPO & DeepSeek-R1 & 45.00 & 50.50 & 18.00 & 42.09 \\
\midrule
\midrule
\multicolumn{7}{l}{\textbf{HotpotQA Dataset}} \\
\midrule
Local Search & & Qwen2.5 72B & \textbf{59.25} & 10.63 & 6.19 & 38.22 \\
Global Search & Qwen2.5 72B & Qwen2.5 72B & 18.49 & 41.88 & 6.19 & 19.78 \\
Drift Search & Qwen2.5 72B & Qwen2.5 72B & 39.62 & 30.00 & 22.38 & 33.89 \\
\rowcolor{mygray}
Deep GraphRAG & Qwen2.5 72B & Qwen2.5 72B & 49.06 & 56.25 & \textbf{24.76} & \textbf{44.67} \\
\rowcolor{mygray}
Deep GraphRAG & Qwen2.5 1.5B & Qwen2.5 72B & 18.87 & 15.00 & 2.86 & 15.97 \\
\rowcolor{mygray}
Deep GraphRAG & Qwen2.5 1.5B-DW GRPO & Qwen2.5 72B & 39.62 & \textbf{57.50} & 20.95 & 38.44 \\
\midrule
Local Search & & DeepSeek-R1 & \textbf{60.94} & 10.00 & 14.76 & 41.11 \\
Global Search & DeepSeek-R1 & DeepSeek-R1 & 20.00 & 48.75 & 9.05 & 22.56 \\
Drift Search & Qwen2.5 72B & DeepSeek-R1 & 39.62 & 38.75 & 22.86 & 35.56 \\
\rowcolor{mygray}
Deep GraphRAG & Qwen2.5 72B & DeepSeek-R1 & 50.38 & 56.25 & \textbf{24.76} & \textbf{45.44} \\
\rowcolor{mygray}
Deep GraphRAG & Qwen2.5 1.5B & DeepSeek-R1 & 20.00 & 18.13 & 3.81 & 17.48 \\
\rowcolor{mygray}
Deep GraphRAG & Qwen2.5 1.5B-DW GRPO & DeepSeek-R1 & 39.62 & 58.13 & 20.95 & 38.56 \\
\bottomrule
\end{tabular}%
} 
\end{table*}

\begin{figure}[H]
\centering
\includegraphics[width=0.85\columnwidth]{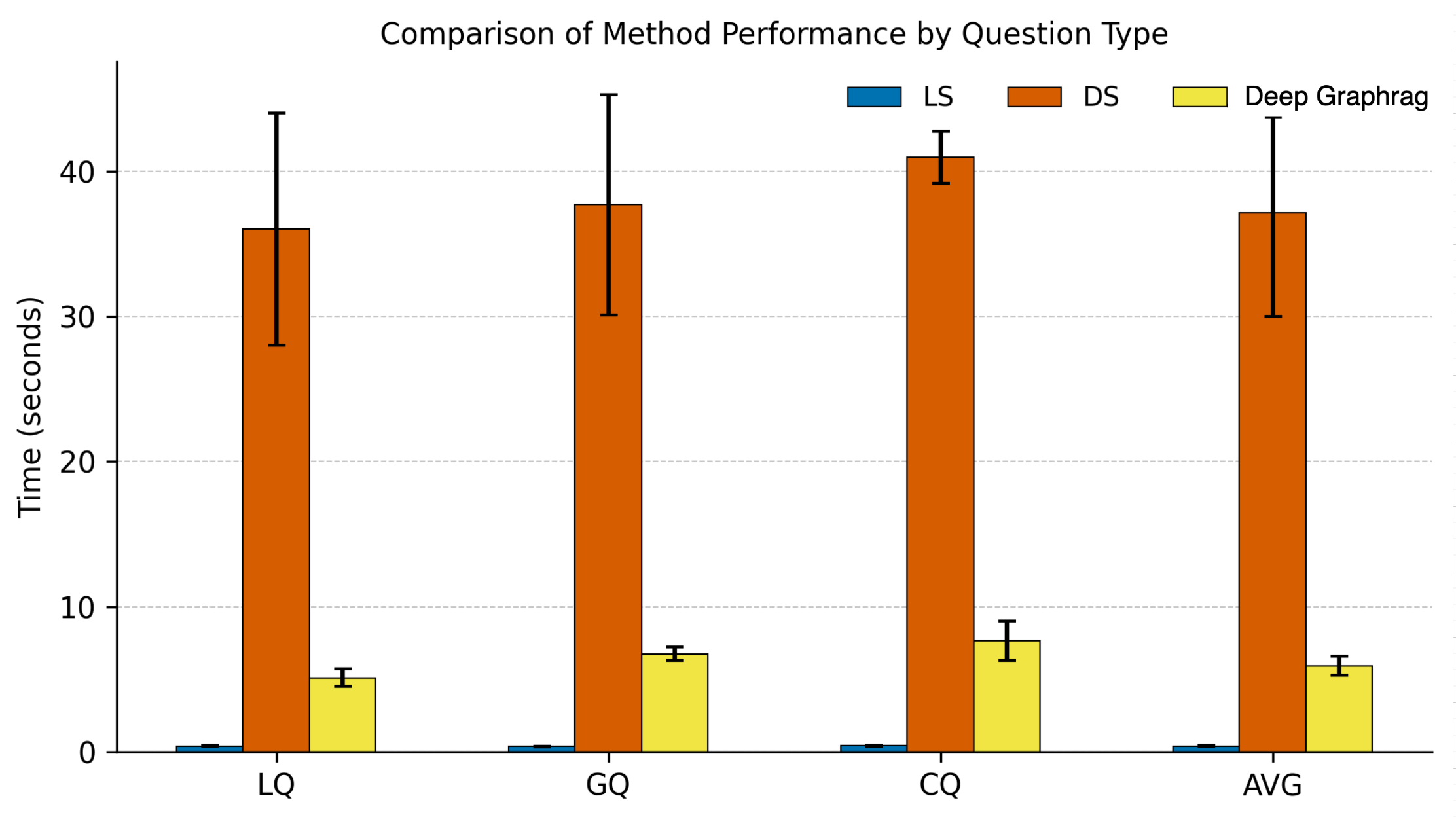}
\caption{Comparison of system processing time (latency) among Local Search (LS), DRIFT Search (DS), and Deep GraphRAG methods on the NQ dataset.}
\label{fig3}
\end{figure}

\section{Results}

\subsection{Performance and Efficiency Analysis}

We conducted a comprehensive evaluation of retrieval performance (Table \ref{tab:performance_combined}) and system latency (Figure \ref{fig3}). Our analysis yields three critical observations.

\textbf{Superiority in Global Reasoning Drives Overall Performance.}
Deep GraphRAG (with a 72B integrator) establishes new state-of-the-art performance, achieving the highest \textbf{EM-Total} scores on both NQ (e.g., 44.69\%) and HotpotQA (e.g., 45.44\%). This overall superiority is primarily driven by a significant advantage in \textbf{Global Questions (GQ)}. On HotpotQA, which demands multi-hop reasoning, our method (e.g., 56.25\%) drastically outperforms Local Search (10.00\%) and the strong Drift Search baseline (38.75\%). This confirms that our hierarchical global-to-local search effectively balances exploration and exploitation to handle queries requiring broad contextual aggregation.

\textbf{Nuanced Trade-offs in Comprehensive Queries.}
Performance on \textbf{Comprehensive Questions (CQ)} is more nuanced. While Deep GraphRAG remains competitive and often wins (e.g., 24.76\% on HotpotQA), it does not universally outperform all baselines. For instance, on NQ with DeepSeek-R1, Local Search achieves a higher EM-CQ (23.20\%) than Deep GraphRAG (19.60\%). This suggests a potential trade-off: our method's strength in hierarchical summarization may, in some cases, obscure the fine-grained local facts needed for certain CQ tasks, an area for future refinement.

\textbf{High-Efficiency Distillation via DW-GRPO.} 
The compact \texttt{Qwen2.5 1.5B-DW GRPO} achieves \textbf{over 94\%} of the 72B model's NQ performance (42.36\%), overcoming the baseline's ``lost-in-the-middle'' issue. By reinforcing information filtering and query alignment, our dynamic weighting strategy enhances contextual robustness, enabling lightweight models to handle complex reasoning.

\textbf{Latency Reduction.}
Figure \ref{fig3} shows Deep GraphRAG achieves an \textbf{86\%} and \textbf{81.6\%} reduction in latency over Drift Search on Local and Global NQ questions.

\subsection{Performance of DW-GRPO}

To rigorously evaluate the efficacy of the proposed Dynamic Weighting Reward GRPO (DW-GRPO), we conducted a comparative analysis against the standard GRPO baseline using the Qwen2.5-1.5B model. The experimental protocol involved a two-stage training pipeline on the HotpotQA dataset: initial Supervised Fine-Tuning (SFT) using distillations from the teacher model (Qwen2.5-72B), followed by Reinforcement Learning (RL) optimization. We track both the aggregated reward and the individual components: \textit{Conciseness}, \textit{Faithfulness}, and \textit{Relevance}.

\begin{figure}[H]
\centering
\includegraphics[width=0.9\columnwidth]{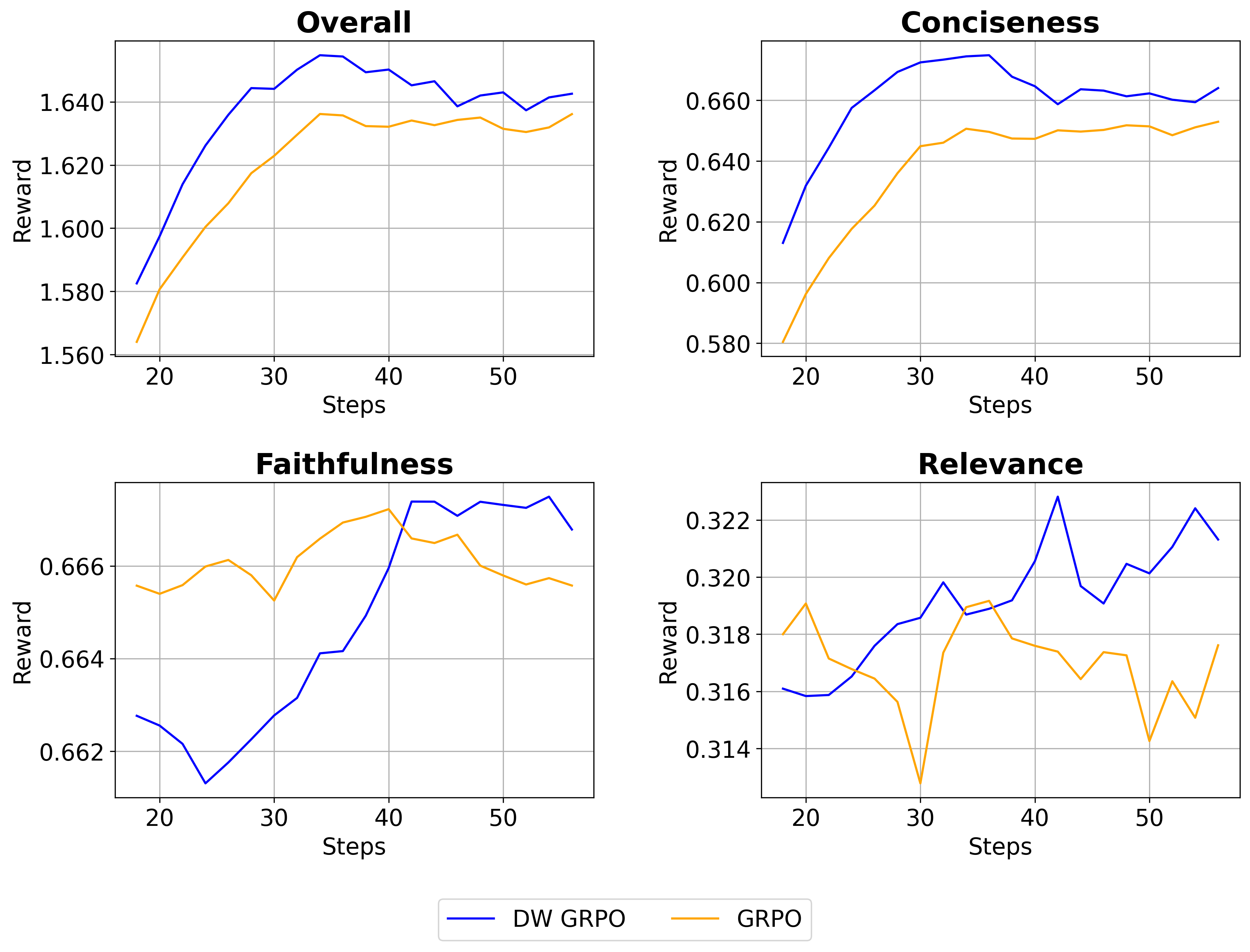}
\caption{Comparison of smoothed reward learning curves between GRPO and DW-GRPO on the test dataset.}
\label{fig4}
\end{figure}

Figure \ref{fig4} illustrates the reward trajectories. The baseline exhibits a marked optimization disparity: while the simple \textit{Conciseness} reward is rapidly maximized, semantic objectives---\textit{Relevance} and \textit{Faithfulness}---stagnate. This exemplifies the ``seesaw effect,'' where the model over-optimizes easy metrics at the expense of complex reasoning.

In contrast, DW-GRPO demonstrates sustained gains across all metrics, validating the adaptive mechanism in Section 2.2.2. By monitoring the reward slope ($\text{slope}_j$), the algorithm detects stalling in \textit{Relevance} and \textit{Faithfulness} and dynamically upweights these lagging components (Eq. 7). This effectively prevents policy collapse into trivial solutions, ensuring robust semantic alignment.

\section{Conclusion}
We introduced Deep GraphRAG, a framework addressing global and complex queries via hierarchical global-to-local retrieval and beam-search re-ranking. Additionally, we proposed DW-GRPO, an adaptive reinforcement learning strategy that enables compact 1.5B models to achieve performance competitive with 72B baselines. Experiments on NQ and HotpotQA demonstrate significant gains in both retrieval accuracy and efficiency. Future work will focus on optimizing the trade-off between global summarization and local fact preservation.

\bibliographystyle{ACM-Reference-Format}

\begin{thebibliography}{23}


\ifx \showCODEN    \undefined \def \showCODEN     #1{\unskip}     \fi
\ifx \showISBNx    \undefined \def \showISBNx     #1{\unskip}     \fi
\ifx \showISBNxiii \undefined \def \showISBNxiii  #1{\unskip}     \fi
\ifx \showISSN     \undefined \def \showISSN      #1{\unskip}     \fi
\ifx \showLCCN     \undefined \def \showLCCN      #1{\unskip}     \fi
\ifx \shownote     \undefined \def \shownote      #1{#1}          \fi
\ifx \showarticletitle \undefined \def \showarticletitle #1{#1}   \fi
\ifx \showURL      \undefined \def \showURL       {\relax}        \fi
\providecommand\bibfield[2]{#2}
\providecommand\bibinfo[2]{#2}
\providecommand\natexlab[1]{#1}
\providecommand\showeprint[2][]{arXiv:#2}

\bibitem[Bruckhaus(2024)]%
        {2406_04369}
\bibfield{author}{\bibinfo{person}{T. Bruckhaus}.} \bibinfo{year}{2024}\natexlab{}.
\newblock \showarticletitle{Rag does not work for enterprises}.
\newblock \bibinfo{journal}{\emph{arXiv preprint arXiv:2406.04369}} (\bibinfo{year}{2024}).
\newblock


\bibitem[Chen et~al\mbox{.}(2023)]%
        {bge_m3}
\bibfield{author}{\bibinfo{person}{Jianlv Chen}, \bibinfo{person}{Shitao Xiao}, \bibinfo{person}{Peitian Zhang}, \bibinfo{person}{Kun Luo}, \bibinfo{person}{Defu Lian}, {and} \bibinfo{person}{Zheng Liu}.} \bibinfo{year}{2023}\natexlab{}.
\newblock \bibinfo{title}{BGE M3-Embedding: Multi-Lingual, Multi-Functionality, Multi-Granularity Text Embeddings Through Self-Knowledge Distillation}.
\newblock
\showeprint[arxiv]{2309.07597}~[cs.CL]


\bibitem[Edge et~al\mbox{.}(2024)]%
        {Edge2024}
\bibfield{author}{\bibinfo{person}{Darren Edge} {et~al\mbox{.}}} \bibinfo{year}{2024}\natexlab{}.
\newblock \showarticletitle{From Local to Global: A Graph RAG Approach to Query-Focused Summarization}.
\newblock \bibinfo{journal}{\emph{arXiv preprint arXiv:2404.16130}} (\bibinfo{year}{2024}).
\newblock


\bibitem[Glass et~al\mbox{.}(2022)]%
        {2207_06300}
\bibfield{author}{\bibinfo{person}{M. Glass}, \bibinfo{person}{G. Rossiello}, \bibinfo{person}{M.~F.~M. Chowdhury}, \bibinfo{person}{A.~R. Naik}, \bibinfo{person}{P. Cai}, {and} \bibinfo{person}{A. Gliozzo}.} \bibinfo{year}{2022}\natexlab{}.
\newblock \showarticletitle{Re2G: Retrieve, rerank, generate}.
\newblock \bibinfo{journal}{\emph{arXiv preprint arXiv:2207.06300}} (\bibinfo{year}{2022}).
\newblock


\bibitem[Gutiérrez et~al\mbox{.}(2025)]%
        {gutiérrez2025hipporagneurobiologicallyinspiredlongterm}
\bibfield{author}{\bibinfo{person}{Bernal~Jiménez Gutiérrez}, \bibinfo{person}{Yiheng Shu}, \bibinfo{person}{Yu Gu}, \bibinfo{person}{Michihiro Yasunaga}, {and} \bibinfo{person}{Yu Su}.} \bibinfo{year}{2025}\natexlab{}.
\newblock \bibinfo{title}{HippoRAG: Neurobiologically Inspired Long-Term Memory for Large Language Models}.
\newblock
\showeprint[arxiv]{2405.14831}~[cs.CL]
\urldef\tempurl%
\url{https://arxiv.org/abs/2405.14831}
\showURL{%
\tempurl}


\bibitem[Han et~al\mbox{.}(2024)]%
        {Han2024}
\bibfield{author}{\bibinfo{person}{H. Han} {et~al\mbox{.}}} \bibinfo{year}{2024}\natexlab{}.
\newblock \showarticletitle{Retrieval-augmented generation with graphs (graphrag)}.
\newblock \bibinfo{journal}{\emph{arXiv preprint arXiv:2501.00309}} (\bibinfo{year}{2024}).
\newblock


\bibitem[Kwiatkowski et~al\mbox{.}(2019)]%
        {10.1162/tacl_a_00276}
\bibfield{author}{\bibinfo{person}{Tom Kwiatkowski}, \bibinfo{person}{Jennimaria Palomaki}, \bibinfo{person}{Olivia Redfield}, \bibinfo{person}{Michael Collins}, \bibinfo{person}{Ankur Parikh}, \bibinfo{person}{Chris Alberti}, \bibinfo{person}{Danielle Epstein}, \bibinfo{person}{Illia Polosukhin}, \bibinfo{person}{Jacob Devlin}, \bibinfo{person}{Kenton Lee}, \bibinfo{person}{Kristina Toutanova}, \bibinfo{person}{Llion Jones}, \bibinfo{person}{Matthew Kelcey}, \bibinfo{person}{Ming-Wei Chang}, \bibinfo{person}{Andrew~M. Dai}, \bibinfo{person}{Jakob Uszkoreit}, \bibinfo{person}{Quoc Le}, {and} \bibinfo{person}{Slav Petrov}.} \bibinfo{year}{2019}\natexlab{}.
\newblock \showarticletitle{Natural Questions: A Benchmark for Question Answering Research}.
\newblock \bibinfo{journal}{\emph{Transactions of the Association for Computational Linguistics}}  \bibinfo{volume}{7} (\bibinfo{year}{2019}), \bibinfo{pages}{453--466}.
\newblock
\showISSN{2307-387X}
\showeprint{https://direct.mit.edu/tacl/article-pdf/doi/10.1162/tacl\_a\_00276/1923288/tacl\_a\_00276.pdf}
\href{https://doi.org/10.1162/tacl_a_00276}{doi:\nolinkurl{10.1162/tacl_a_00276}}


\bibitem[Lewis et~al\mbox{.}(2020)]%
        {lewis2020retrieval}
\bibfield{author}{\bibinfo{person}{Patrick Lewis}, \bibinfo{person}{Ethan Perez}, \bibinfo{person}{Aleksandra Piktus}, \bibinfo{person}{Fabio Petroni}, \bibinfo{person}{Vladimir Karpukhin}, \bibinfo{person}{Naman Goyal}, \bibinfo{person}{Heinrich K{\"u}ttler}, \bibinfo{person}{Mike Lewis}, \bibinfo{person}{Wen-tau Yih}, \bibinfo{person}{Tim Rockt{\"a}schel}, \bibinfo{person}{Sebastian Riedel}, {and} \bibinfo{person}{Douwe Kiela}.} \bibinfo{year}{2020}\natexlab{}.
\newblock \showarticletitle{Retrieval-Augmented Generation for Knowledge-Intensive NLP Tasks}. In \bibinfo{booktitle}{\emph{Advances in Neural Information Processing Systems}}, \bibfield{editor}{\bibinfo{person}{H.~Larochelle}, \bibinfo{person}{M.~Ranzato}, \bibinfo{person}{R.~Hadsell}, \bibinfo{person}{M.~F. Balcan}, {and} \bibinfo{person}{H.~Lin}} (Eds.), Vol.~\bibinfo{volume}{33}. \bibinfo{publisher}{Curran Associates, Inc.}, \bibinfo{pages}{9459--9474}.
\newblock


\bibitem[Li et~al\mbox{.}(2024b)]%
        {li2024graphreaderbuildinggraphbasedagent}
\bibfield{author}{\bibinfo{person}{Shilong Li}, \bibinfo{person}{Yancheng He}, \bibinfo{person}{Hangyu Guo}, \bibinfo{person}{Xingyuan Bu}, \bibinfo{person}{Ge Bai}, \bibinfo{person}{Jie Liu}, \bibinfo{person}{Jiaheng Liu}, \bibinfo{person}{Xingwei Qu}, \bibinfo{person}{Yangguang Li}, \bibinfo{person}{Wanli Ouyang}, \bibinfo{person}{Wenbo Su}, {and} \bibinfo{person}{Bo Zheng}.} \bibinfo{year}{2024}\natexlab{b}.
\newblock \bibinfo{title}{GraphReader: Building Graph-based Agent to Enhance Long-Context Abilities of Large Language Models}.
\newblock
\showeprint[arxiv]{2406.14550}~[cs.CL]
\urldef\tempurl%
\url{https://arxiv.org/abs/2406.14550}
\showURL{%
\tempurl}


\bibitem[Li et~al\mbox{.}(2025)]%
        {li2025rglgraphcentricmodularframework}
\bibfield{author}{\bibinfo{person}{Yuan Li}, \bibinfo{person}{Jun Hu}, \bibinfo{person}{Jiaxin Jiang}, \bibinfo{person}{Zemin Liu}, \bibinfo{person}{Bryan Hooi}, {and} \bibinfo{person}{Bingsheng He}.} \bibinfo{year}{2025}\natexlab{}.
\newblock \bibinfo{title}{RGL: A Graph-Centric, Modular Framework for Efficient Retrieval-Augmented Generation on Graphs}.
\newblock
\showeprint[arxiv]{2503.19314}~[cs.IR]
\urldef\tempurl%
\url{https://arxiv.org/abs/2503.19314}
\showURL{%
\tempurl}


\bibitem[Li et~al\mbox{.}(2024a)]%
        {2410_08815}
\bibfield{author}{\bibinfo{person}{Z. Li}, \bibinfo{person}{X. Chen}, \bibinfo{person}{H. Yu}, \bibinfo{person}{H. Lin}, \bibinfo{person}{Y. Lu}, \bibinfo{person}{Q. Tang}, \bibinfo{person}{F. Huang}, \bibinfo{person}{X. Han}, \bibinfo{person}{L. Sun}, {and} \bibinfo{person}{Y. Li}.} \bibinfo{year}{2024}\natexlab{a}.
\newblock \showarticletitle{Structrag: Boosting knowledge intensive reasoning of llms via inference-time hybrid information structurization}.
\newblock \bibinfo{journal}{\emph{arXiv preprint arXiv:2410.08815}} (\bibinfo{year}{2024}).
\newblock


\bibitem[Ling et~al\mbox{.}(2023)]%
        {2305_18703}
\bibfield{author}{\bibinfo{person}{C. Ling}, \bibinfo{person}{X. Zhao}, \bibinfo{person}{J. Lu}, \bibinfo{person}{C. Deng}, \bibinfo{person}{C. Zheng}, \bibinfo{person}{J. Wang}, \bibinfo{person}{T. Chowdhury}, \bibinfo{person}{Y. Li}, \bibinfo{person}{H. Cui}, {and} \bibinfo{person}{X.~Zhang et al.}} \bibinfo{year}{2023}\natexlab{}.
\newblock \showarticletitle{Domain specialization as the key to make large language models disruptive: A comprehensive survey}.
\newblock \bibinfo{journal}{\emph{arXiv preprint arXiv:2305.18703}} (\bibinfo{year}{2023}).
\newblock


\bibitem[Luo et~al\mbox{.}(2025)]%
        {luo2025gfmraggraphfoundationmodel}
\bibfield{author}{\bibinfo{person}{Linhao Luo}, \bibinfo{person}{Zicheng Zhao}, \bibinfo{person}{Gholamreza Haffari}, \bibinfo{person}{Dinh Phung}, \bibinfo{person}{Chen Gong}, {and} \bibinfo{person}{Shirui Pan}.} \bibinfo{year}{2025}\natexlab{}.
\newblock \bibinfo{title}{GFM-RAG: Graph Foundation Model for Retrieval Augmented Generation}.
\newblock
\showeprint[arxiv]{2502.01113}~[cs.IR]
\urldef\tempurl%
\url{https://arxiv.org/abs/2502.01113}
\showURL{%
\tempurl}


\bibitem[Rafailov et~al\mbox{.}(2023)]%
        {rafailov2023direct}
\bibfield{author}{\bibinfo{person}{Rafael Rafailov}, \bibinfo{person}{Archit Sharma}, \bibinfo{person}{Eric Mitchell}, \bibinfo{person}{Christopher~D Manning}, \bibinfo{person}{Stefano Ermon}, {and} \bibinfo{person}{Chelsea Finn}.} \bibinfo{year}{2023}\natexlab{}.
\newblock \showarticletitle{Direct preference optimization: Your language model is secretly a reward model}.
\newblock \bibinfo{journal}{\emph{Advances in neural information processing systems}}  \bibinfo{volume}{36} (\bibinfo{year}{2023}), \bibinfo{pages}{53728--53741}.
\newblock


\bibitem[Research(2024)]%
        {Microsoft2024}
\bibfield{author}{\bibinfo{person}{Microsoft Research}.} \bibinfo{year}{2024}\natexlab{}.
\newblock \showarticletitle{Introducing Drift Search: Combining Global and Local Search Methods to Improve Quality and Efficiency}.
\newblock \bibinfo{journal}{\emph{Microsoft Research Blog}} (\bibinfo{year}{2024}).
\newblock
\urldef\tempurl%
\url{https://www.microsoft.com/en-us/research/blog/introducing-drift-search-combining-global-and-local-search-methods-to-improve-quality-and-efficiency/}
\showURL{%
\tempurl}


\bibitem[Sarthi et~al\mbox{.}(2024)]%
        {2401_18059}
\bibfield{author}{\bibinfo{person}{P. Sarthi}, \bibinfo{person}{S. Abdullah}, \bibinfo{person}{A. Tuli}, \bibinfo{person}{S. Khanna}, \bibinfo{person}{A. Goldie}, {and} \bibinfo{person}{C.~D. Manning}.} \bibinfo{year}{2024}\natexlab{}.
\newblock \showarticletitle{Raptor: Recursive abstractive processing for tree-organized retrieval}.
\newblock \bibinfo{journal}{\emph{arXiv preprint arXiv:2401.18059}} (\bibinfo{year}{2024}).
\newblock


\bibitem[Schulman et~al\mbox{.}(2015)]%
        {schulman2015trust}
\bibfield{author}{\bibinfo{person}{John Schulman}, \bibinfo{person}{Sergey Levine}, \bibinfo{person}{Pieter Abbeel}, \bibinfo{person}{Michael Jordan}, {and} \bibinfo{person}{Philipp Moritz}.} \bibinfo{year}{2015}\natexlab{}.
\newblock \showarticletitle{Trust region policy optimization}. In \bibinfo{booktitle}{\emph{International conference on machine learning}}. PMLR, \bibinfo{pages}{1889--1897}.
\newblock


\bibitem[Schulman et~al\mbox{.}(2017)]%
        {schulman2017proximal}
\bibfield{author}{\bibinfo{person}{John Schulman}, \bibinfo{person}{Filip Wolski}, \bibinfo{person}{Prafulla Dhariwal}, \bibinfo{person}{Alec Radford}, {and} \bibinfo{person}{Oleg Klimov}.} \bibinfo{year}{2017}\natexlab{}.
\newblock \showarticletitle{Proximal policy optimization algorithms}.
\newblock \bibinfo{journal}{\emph{arXiv preprint arXiv:1707.06347}} (\bibinfo{year}{2017}).
\newblock


\bibitem[Shao et~al\mbox{.}(2024)]%
        {shao2024deepseekmath}
\bibfield{author}{\bibinfo{person}{Zhihong Shao}, \bibinfo{person}{Peiyi Wang}, \bibinfo{person}{Qihao Zhu}, \bibinfo{person}{Runxin Xu}, \bibinfo{person}{Junxiao Song}, \bibinfo{person}{Xiao Bi}, \bibinfo{person}{Haowei Zhang}, \bibinfo{person}{Mingchuan Zhang}, \bibinfo{person}{YK Li}, \bibinfo{person}{Yang Wu}, {et~al\mbox{.}}} \bibinfo{year}{2024}\natexlab{}.
\newblock \showarticletitle{Deepseekmath: Pushing the limits of mathematical reasoning in open language models}.
\newblock \bibinfo{journal}{\emph{arXiv preprint arXiv:2402.03300}} (\bibinfo{year}{2024}).
\newblock


\bibitem[Tang et~al\mbox{.}(2020)]%
        {10.1145/3383313.3412236}
\bibfield{author}{\bibinfo{person}{Hongyan Tang}, \bibinfo{person}{Junning Liu}, \bibinfo{person}{Ming Zhao}, {and} \bibinfo{person}{Xudong Gong}.} \bibinfo{year}{2020}\natexlab{}.
\newblock \showarticletitle{Progressive Layered Extraction (PLE): A Novel Multi-Task Learning (MTL) Model for Personalized Recommendations}. In \bibinfo{booktitle}{\emph{Proceedings of the 14th ACM Conference on Recommender Systems}} (Virtual Event, Brazil) \emph{(\bibinfo{series}{RecSys '20})}. \bibinfo{publisher}{Association for Computing Machinery}, \bibinfo{address}{New York, NY, USA}, \bibinfo{pages}{269–278}.
\newblock
\showISBNx{9781450375832}
\href{https://doi.org/10.1145/3383313.3412236}{doi:\nolinkurl{10.1145/3383313.3412236}}


\bibitem[Tang et~al\mbox{.}(2024)]%
        {2412_01572}
\bibfield{author}{\bibinfo{person}{X. Tang}, \bibinfo{person}{Q. Gao}, \bibinfo{person}{J. Li}, \bibinfo{person}{N. Du}, \bibinfo{person}{Q. Li}, {and} \bibinfo{person}{S. Xie}.} \bibinfo{year}{2024}\natexlab{}.
\newblock \showarticletitle{Mba-rag: a bandit approach for adaptive retrieval-augmented generation through question complexity}.
\newblock \bibinfo{journal}{\emph{arXiv preprint arXiv:2412.01572}} (\bibinfo{year}{2024}).
\newblock


\bibitem[Xu et~al\mbox{.}(2023)]%
        {2310_04408}
\bibfield{author}{\bibinfo{person}{F. Xu}, \bibinfo{person}{W. Shi}, {and} \bibinfo{person}{E. Choi}.} \bibinfo{year}{2023}\natexlab{}.
\newblock \showarticletitle{Recomp: Improving retrieval-augmented lms with compression and selective augmentation}.
\newblock \bibinfo{journal}{\emph{arXiv preprint arXiv:2310.04408}} (\bibinfo{year}{2023}).
\newblock


\bibitem[Yang et~al\mbox{.}(2018)]%
        {yang2018hotpotqadatasetdiverseexplainable}
\bibfield{author}{\bibinfo{person}{Zhilin Yang}, \bibinfo{person}{Peng Qi}, \bibinfo{person}{Saizheng Zhang}, \bibinfo{person}{Yoshua Bengio}, \bibinfo{person}{William~W. Cohen}, \bibinfo{person}{Ruslan Salakhutdinov}, {and} \bibinfo{person}{Christopher~D. Manning}.} \bibinfo{year}{2018}\natexlab{}.
\newblock \bibinfo{title}{HotpotQA: A Dataset for Diverse, Explainable Multi-hop Question Answering}.
\newblock
\showeprint[arxiv]{1809.09600}~[cs.CL]
\urldef\tempurl%
\url{https://arxiv.org/abs/1809.09600}
\showURL{%
\tempurl}


\end{thebibliography}

% ...

\end{document}